\documentclass[aps,prb,reprint]{revtex4-2}
\usepackage{amsmath,amssymb} 
\usepackage{amsmath} 
\usepackage{amsfonts} 
\usepackage{bm} 
\usepackage[dvipdfmx]{graphicx} 
\usepackage{braket} 
\usepackage{physics}
\usepackage{color}
\usepackage{comment}
\usepackage{mathtools}
\usepackage[normalem]{ulem}

\begin{document}
\title{Pinch-point spectral singularity from the interference of topological loop states}

\author{Masafumi Udagawa$^{1}$}
\email{Masafumi.Udagawa@gakushuin.ac.jp}
\author{Hiroki Nakai$^2$}
\email{nakai-hiroki3510@g.ecc.u-tokyo.ac.jp}
\author{Chisa Hotta$^2$}
\email{chisa@phys.c.u-tokyo.ac.jp}
\affiliation{$^1$Department of Physics, Gakushuin University, Mejiro, Tosshima-ku, Tokyo 171-8588, Japan}
\affiliation{$^2$Graduate School of Arts and Sciences, University of Tokyo, Meguro-ku, Tokyo 153-8902, Japan}

\date{\today}

\begin{abstract} 
Pinch point is a spectral discontinuity found in the neutron diffraction image of spin ice.
Similar spectral singularity is commonly observed in a broad range of systems that 
have a close connection with flat bands. 
We focus on the electron flat band and its two topologically distinct classes of wavefunction: 
the compact localized state (CLS), and the non-contractible loop state (NLS). 
We establish their simple mathematical relationship, 
showing that different Bloch NLSs can be derived as momentum derivatives of a Bloch CLS, 
depending on the approaching direction toward the singular point. 
This CLS-NLS correspondence helps visualize the pinch point as an interference pattern among NLSs through a ``polarizer", 
which encodes the information about the location of singular momentum and the experimental techniques 
like spin-polarized photoemission spectroscopy. 
It helps extract topological information knit to microscopic electronic and magnetic structures. 
\end{abstract} 

\maketitle
{\it Introduction.} 
Characterization of quantum states using the language of topology often encounters difficulty 
in confirming it experimentally via local observables. 
One of the few exceptions to date is the integer quantum Hall systems~\cite{Klitzing1986} whose Hall conductivity is quantized in units of $e^2/h$. In this light, the global loop states of flat bands may offer another example because 
they should have strong relevance to the observable singularity of the dynamical structure factor called pinch points. 
\par
The pinch point originally drew attention as a singularity of the magnetic structure factor of spin ice compounds, 
Ho$_2$Ti$_2$O$_7$~\cite{Fennell2009,Kadowaki2009}. 
The spin ice state consists of a macroscopic number of classical states that satisfy 
a local constraint known as the ice rule, wherein the spin configuration 
within a tetrahedron sums to zero~\cite{Udagawa2021, Bramwell2001}. 
From a field-theory point of view, the ice rule is a divergence-free condition in analogy 
with the divergence-free magnetic field lines in electromagnetism, 
which produces the algebraic dipolar spin correlation~\cite{Huse2003} 
and a resultant pinch point singularity at $\vb*{k}^*=0$~\cite{Isakov2004, Henley2005}. 
However, pinch point is not a unique concept of spin ice. 
\par
The common origin of pinch points underlying a variety of systems is a flat band. 
It is known that the magnetic correlation of spin ice is accurately 
described within a large-${\cal N}$ approximation, 
which attributes the magnetic structure factor of spin ice to the flat band eigenfunctions in momentum space
~\cite{Isakov2004,Garanin1999,Henley2005,Sen2013}. 
This description offers two perspectives. 
Firstly, it highlights the ubiquity of pinch points in various systems ranging from frustrated magnets 
hosting classical spin liquids~\cite{Moessner1998,Garanin1999,Canals2001,Rehn2016,Benton2021}, 
tensor spin liquids~\cite{Abhinav2018}, water ice~\cite{Li1994,Isakov2015}, ferroelectric models~\cite{Youngblood1980}, 
to the materials with electronic flat bands
~\cite{Yan2023-ele,Xie2019,Mao2020,Kang2020_FeSn,Kang2020_CoSn,Lin2018}. 
Secondly, there is a rich mathematical structure underlying the flat band eigenstates
~\cite{Sutherland1986,Bergman2008,Maimaiti2017, Rhim2019,Hwang2021_1,Hwang2021_2,Hwang2021_3,Rhim2020,Rhim2021,
Miyahara2005,Maimaiti2017,Maimaiti2019,Mizoguchi2019,Maimaiti2021,Essafi2017aa,Bilitewski2018}
including global loop states~\cite{Sutherland1986,Bergman2008,Maimaiti2017, Rhim2019}, 
quantum geometry~\cite{Hwang2021_1,Hwang2021_2,Hwang2021_3,Rhim2020}, 
and singularity at band-touching points~\cite{Rhim2021}. 
Despite the knowledge accumulated on flat band and pinch point physics, 
their mutual relationship has not yet been clearly established. 
\par
{\it Flat band phenomena}.\;
Let us illustrate what is known about flat bands using the kagome lattice. 
On the flat-band energy level, there exist states of at least the number of unit cells, $N_c$. 
The absence of momentum dependence implies that these states can be written as 
localized states in real space, typically as the closed loops of finite length called a compact loop state (CLS). 
\par
Suppose that we have a single CLS state positioned at $\vb*{R}_0$, denoted as $|u^{\rm CLS}_{\vb*{R}_0}\rangle$, 
where one can generate its $N_c$ copies at $\{\vb*{R}_l\}$ by translation. 
By superposing them with a phase factor $e^{i\vb*{k}\cdot\vb*{R}_l}$, 
one naturally expects that $N_c$ independent Bloch states constitute a flat band as
\begin{eqnarray}
|u_{\vb*{k}}^{\rm CLS}\rangle\equiv\frac{1}{\sqrt{N_c}}\sum_{l=0}^{N_c-1} e^{i\vb*{k}\cdot\vb*{R}_l}
|u^{\rm CLS}_{\vb*{R}_l}\rangle. 
\label{eq:Bloch}
\end{eqnarray} 
However, in a popular example of a tight-binding model on a kagome lattice, 
they fail to span a complete basis 
because $\sum_{l=0}^{N_c-1} |u^{\rm CLS}_{\vb*{R}_l}\rangle=0$ (non-local constraint), 
resulting in only $N_c-1$ independent Bloch states in Eq.~(\ref{eq:Bloch}). 
The missing state is located at a momentum $\vb*{k}^*=0$. 
\par
Bergmann {\it et al.}~\cite{Bergman2008} identified these missing states as non-contractible loop states (NLS): 
a global loop state winding around the system, topologically different from CLS. 
The kagome lattice has two NLSs extending in different directions, 
and one of them complements the missing flat band state, whereas the other belongs to the dispersive band. 
The presence of two NLS ensures that the dispersive band touches the flat band at $\vb*{k}^*=0$. 
\par
Indeed, the following facts on the ``singular momentum" $\vb*{k^*}$ have been recursively discussed~\cite{Yan2023-sl}; 
(i) vanishing norm of the CLS $\langle u^{\rm CLS}_{\vb*{k}^*}|u^{\rm CLS}_{\vb*{k}^*}\rangle=0$, 
(ii) existence of NLSs, 
(iii) the discontinuity of flat band wave functions, CLS ($\vb*{k}\ne \vb*{k^*}$) and NLS($\vb*{k}=\vb*{k^*}$), 
(iv) band touching, and (v) pinch point. 
Rhim and Yang referred to (i) - (iii) as different facets of the same physics, 
while they further showed counter-example; (iv) exists but (ii) and (iii) do not~\cite{Rhim2019,Rhim2020}. 
There is a list of how the flat band phenomena like (ii) and (iv) are interpreted within a large-${\cal N}$ scheme 
in the classical spin liquid state of magnets~\cite{Yan2023-sl}. 
Although (i)-(v) are commonly observed in flat band systems, 
their mutual relationships are not necessarily clear. 
\par
This Letter aims at establishing a simple relationship between NLS and CLS and 
make transparent connections among the aforementioned flat band phenomena. 
Through a consistent characterization, 
we can see how the pinch point spectrum arises from NLS interference, 
thereby enabling the extraction of topological information from observable pinch points. 
\par
\begin{figure}[t]
	\begin{center}
	\includegraphics[width=8.5cm]{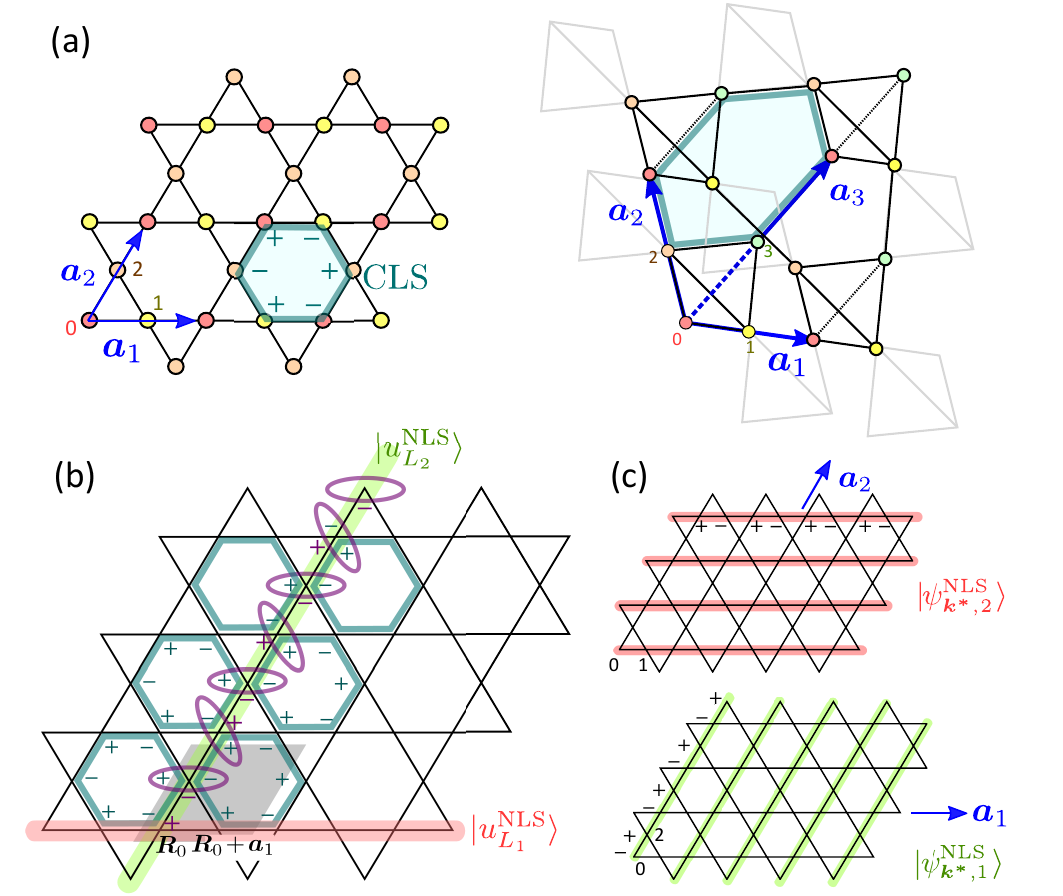}
	\caption{
        (a) Kagome and pyrochlore lattices with unit vectors $\vb*{a}_{\mu}$ and sublattice indices. 
         Hexagonal compact localized states (CLSs) are shown. 
	(b) Non-contractible loop states (NLSs), $|u^{\rm NLS}_{2(1)}\rangle$. 
	(c) Bloch NLSs, $|\psi^{\rm NLS}_{\vb*{k^*},1(2)}\rangle$,
         obtained by superposing the single-loop NLS $|u^{\rm NLS}_{L_{2(1)}}\rangle$ 
	 in the direction $1(2)$.     
                 }
	\label{f1}
	\end{center}
\end{figure}

 
{\it Preliminaries.}\; 
As the simplest platform, we consider a tight-binding Hamiltonian with nearest-neighbor hopping terms, 
$\mathcal{H} = -t\sum_{\langle i,j\rangle}(|i\rangle\langle j| + {\rm H.c.})$, 
on kagome and pyrochlore lattices (Fig.~\ref{f1}(a)) 
for spatial dimensions, $d=2$ and $3$, and number of sublattices, $n_s=3$ and $4$, respectively. 
Let the lattice vector be given as 
$\vb*{R}_{\:l}=\sum_{\mu=1}^d l_\mu \vb*{a}_\mu,\;$ $l_\mu \in {\mathbb{Z}}$ 
with a unit vector $\vb*{a}_\mu$ in the $\mu$-th direction, 
along which we impose periodic boundary conditions. 
The reciprocal lattice vectors $\vb*{b}_\eta$ satisfy $\vb*{a}_\mu\cdot \vb*{b}_\eta=2\pi \delta_{\mu\eta}$. 
In kagome and pyrochlore lattices, sublattice vectors can be chosen as 
$\vb*{r}_0=0$ and $\vb*{r}_\mu = \vb*{a}_{\mu}/2$ for $\mu=1,\cdots,n_s-1$. 
\par
{\it CLS-NLS correspondence.}\; 
Let us continue with the kagome lattice. 
We may choose a primitive CLS around a hexagon as (Fig.~\ref{f1}(a)),
\begin{align}
|u^{\rm CLS}_{\vb*{R}_l}\rangle &= (|\vb*{R}_l, 1\rangle - |\vb*{R}_l, 2\rangle) 
+ (|\vb*{R}_l+ \vb*{a}_1, 2\rangle - |\vb*{R}_l + \vb*{a}_1, 0\rangle) \nonumber\\	
&+ (|\vb*{R}_l + \vb*{a}_2, 0\rangle - |\vb*{R}_l + \vb*{a}_2, 1\rangle), 
\label{eq:loop6}
\end{align}
where $|\vb*{R}_l, \mu \rangle$ denotes 
the electron state on the $\mu$-th sublattice of the $l$-th unit cell. 
Similarly, we can write down a single global NLS $L_{\mu}$ as
\begin{equation}
|u^{\rm NLS}_{L_{\mu}}\rangle = \sum_{l\in {\rm L}_{\mu}}(|\vb*{R}_l, 0\rangle - |\vb*{R}_l, \mu\rangle),
\label{eq:nls_wavefunction}
\end{equation}
winding around a global loop $L_{1(2)}$ in the direction 1(2)(Fig.~\ref{f1}(b)).
Both CLS and NLS wavefunctions explicitly satisfy the divergence-free condition~\cite{Henley2010}, 
i.e. the coefficients of $|\vb*{R}_l, \mu\rangle$ sum to zero in each unit cell. 
\par
Once we choose an elementary CLS, 
its translation and superposition will provide a set of Bloch CLS, 
\begin{eqnarray}
&& |u^{\rm CLS}_{\vb*{k}} \rangle = \frac{1}{\sqrt{N_c}}\sum_{l=0}^{N_c-1}
e^{i\vb*{k}\cdot\vb*{R}_l}
\Big( \sum_{\mu=0}^{n_s}  v_{\mu,\bm k} |\vb*{R}_l, \mu\rangle\Big), 
\label{eq:CLS_detail}
\\ &&
\left\{
\begin{array}{l}
v_{0,\vb*{k}} = e^{-ik_2} - e^{-ik_1}, \\
v_{1,\vb*{k}} = 1 - e^{-ik_2},\\
v_{2,\vb*{k}} = e^{-ik_1} - 1.
\end{array}
\right. 
\label{eq:bloch-v0}
\end{eqnarray}
In determining the vector, 
$\vb*{v}_{\vb*{k}} = (v_{0,\vb*{k}}, v_{1,\vb*{k}}, v_{2,\vb*{k}})$, 
the overlap of neighboring $|u^{\rm CLS}_{\vb*{R}_l} \rangle$'s, 
each spreading over a few unit cells, is essential. 
For example, $|\vb*{R}_l+ \vb*{a}_1, 2\rangle$ is shared by two CLSs, 
$|u^{\rm CLS}_{\vb*{R}_l}\rangle$ and $|u^{\rm CLS}_{\vb*{R}_l+ \vb*{a}_1}\rangle$, 
and in the Bloch state $|u^{\rm CLS}_{\vb*{k}} \rangle$, 
the former gives $e^{i\vb*{k}\cdot(\vb*{R}_l+ \vb*{a}_1)} e^{-ik_1} |\vb*{R}_l+ \vb*{a}_1, 2\rangle$, while 
the latter, $-e^{i\vb*{k}\cdot(\vb*{R}_l+ \vb*{a}_1)}|\vb*{R}_l+ \vb*{a}_1, 2\rangle$. 
Their sum yields $v_{2,\vb*{k}}$ in Eq.~(\ref{eq:bloch-v0}), where the factor $e^{-ik_1}$ appears 
because CLSs are shared in the direction 1.
$\vb*{v}_{\vb*{k}}$ is also concerned with the norm of the Bloch CLS: 
$(\langle u^{\rm CLS}_{\vb*{k}}|u^{\rm CLS}_{\vb*{k}}\rangle)^{\frac{1}{2}}$ 
$=|\vb*{v}_{\vb*{k}}|=2 [\sin^2\frac{k_2-k_1}{2} + \sin^2\frac{k_2}{2} + \sin^2\frac{k_1}{2}]^{1/2}$. 
Then, $|\vb*{v}_{\vb*{k}}|=0$ at $\vb*{k}=0$, where $|u^{\rm CLS}_{\vb*{k}} \rangle$ cannot be defined, 
i.e. $\vb*{k}= \vb*{k^*}=0$ gives a singular momentum. 
\par
We are now ready to introduce the CLS-NLS correspondence. 
The first observation on Eq.~(\ref{eq:CLS_detail}) is that 
if we replace the coefficient $v_{\mu,\bm k}$ with one of its momentum derivatives, $\frac{\partial v_{\mu,\vb*{k}}}{\partial k_{\eta}}$, 
and take $\vb*{k}\to \vb*{k^*}$, we find for example, 
\begin{align}
|\psi^{\rm NLS}_{\vb*{k}^*,1}\rangle&\equiv\frac{1}{\sqrt{N_c}}\sum_{l=0}^{N_c-1}
e^{i\vb*{k}^*\cdot\vb*{R}_l}
\sum_{\mu=0}^{n_s-1}  \frac{\partial v_{\mu,\vb*{k}^*}}{\partial k_1} |\vb*{R}_l, \mu\rangle
\nonumber\\
&=
\frac{i}{\sqrt{N_c}}\sum_{l=0}^{N_c-1}( |\vb*{R}_l, 0\rangle - |\vb*{R}_l, 2\rangle),
\label{eq:bloch_derivative}
\end{align}
which is nothing but a Bloch NLS, i.e. the equal-weight superposition of the one-loop NLS 
$|u^{\rm NLS}_{L_2} \rangle$ (Eq.~(\ref{eq:nls_wavefunction})) translated one by one in the direction $1$.
The factor $\frac{\partial v_{\mu,\vb*{k}^*}}{\partial k_1}$ is proportional to the weight of NLS state at sublattice $\mu$.
Similarly, the $k_2$ derivative results in another Bloch NLS $|\psi^{\rm NLS}_{\vb*{k}^*,2}\rangle$, which superposes another one-loop NLS, $|u^{\rm NLS}_{L_1} \rangle$, translated in the direction $2$. 
To see why the $k$-derivative can extract NLS from CLS, let us look at their relationships in Fig.~\ref{f1}(b). 
Along the one-loop NLS, $|u^{\rm NLS}_{L_{2}}\rangle$, 
each site belonging to either $\mu=0$ or $2$ hinges two adjacent CLSs as marked with ovals. 
Acting on the phase factor $e^{-ik_1}$ assigned to these hinge sites, the $k$-derivative extracts the one-loop NLS $|u^{\rm NLS}_{L_2} \rangle$. 
\par
The second observation is that such $k$-derivative naturally appears in the process of taking the limit, 
$\vb*{k}\to\vb*{k}^*$. 
To see this, we introduce the normalized Bloch CLS, 
\begin{equation}
  |\psi^{\rm CLS}_{\vb*{k}} \rangle =
\frac{1}{\sqrt{N_c}}\sum_{l=0}^{N_c-1}
e^{i\vb*{k}\cdot\vb*{R}_l}
\sum_{\mu=0}^{n_s}  \frac{v_{\mu,\bm k}}{|\vb*{v}_{\vb*{k}}|} |\vb*{R}_l, \mu\rangle.
\label{eq:normalizedCLS_detail}
\end{equation}
In approaching the singular point as $\vb*{k} = \vb*{k}^* + \delta k_{\eta}\vb*{b}_{\eta}$ and $\delta k_{\eta}\to0$, 
the denominator vanishes linearly as $|\vb*{v}_{\vb*{k}}|\to C\delta k_{\eta}$. 
Given that $\vb*{v}_{\vb*{k}^*}=0$, the coefficient of Eq.~(\ref{eq:normalizedCLS_detail}) approaches
\begin{eqnarray}
\frac{v_{\mu,\bm k}}{|\vb*{v}_{\vb*{k}}|}\to 
\frac{v_{\mu,\vb*{k}^* + \delta k_{\eta}\vb*{b}_{\eta}} -  v_{\mu,\vb*{k}^*}}{C\delta k_{\eta}}
\propto\frac{\partial v_{\mu,\vb*{k}^*}}{\partial k_{\eta}}.
\label{eq:normalizedCLS_coef}
\end{eqnarray}
Combing Eqs.~(\ref{eq:bloch_derivative}) and (\ref{eq:normalizedCLS_coef}), 
it is naturally shown that the Bloch NLS $|\psi^{\rm NLS}_{\vb*{k}^*,\eta}\rangle$ is obtained 
as the $\vb*{k}\to\vb*{k}^*$ limit of normalized Bloch CLS. 
Because Eq.~(\ref{eq:normalizedCLS_coef}) means that 
different NLS state appears depending on the direction $\eta$ to approach $\vb*{k}^*$, 
there is a discontinuity of flat band wavefunction at $\vb*{k}=\vb*{k}^*$. 
\par
One can approach the singular point from an arbitrary direction as 
$\vb*{k}=\vb*{k}^*+(\delta k_1\vb*{b}_{1} + \delta k_2\vb*{b}_{2})/(2\pi)$, 
and the Bloch CLS converges to a linearly combined two Bloch NLSs as 
\begin{equation} 
 |\psi^{\rm CLS}_{\vb*{k}} \rangle\to C_1\delta k_1|\psi^{\rm NLS}_{\vb*{k}^*, 1}\rangle + C_2\delta k_2|\psi^{\rm NLS}_{\vb*{k}^*, 2}\rangle.
\label{eq:NLS_convergence}
\end{equation}
The simplest CLS-NLS correspondence, 
$|\psi^{\rm CLS}_{\vb*{k}}\rangle\rightarrow |\psi^{\rm NLS}_{\vb*{k}^*, 1(2)}\rangle$, 
is attained when approaching $\vb*{k}^*$ along the reciprocal vector, $\vb*{b}_{1(2)}$, 
where we have 
$\frac{\partial}{\partial k_1}\vb*{v}_{\vb*{k}^*}=(\frac{\partial v_{0,\vb*{k}^*}}{\partial k_1}, \frac{\partial v_{1,\vb*{k}^*}}{\partial k_1}, \frac{\partial v_{2,\vb*{k}^*}}{\partial k_1})\propto(+1, 0, -1)$ for $|\psi^{\rm NLS}_{\vb*{k}^*, 1}\rangle$,
and $\frac{\partial}{\partial k_2}\vb*{v}_{\vb*{k}^*}\propto(+1, -1, 0)$ for $|\psi^{\rm NLS}_{\vb*{k}^*, 2}\rangle$. 
They are the weight of $\mu=(0,1,2)$ sublattices constituting the associated one-loop NLS states, 
$|u^{\rm NLS}_{L_{2}}\rangle$ and $|u^{\rm NLS}_{L_{1}}\rangle$, respectively (see Fig.~\ref{f1}(b)). 
\par
There might be a small complication in matching $|\psi^{\rm NLS}_{\vb*{k}^*, 1(2)}\rangle$ with $|u^{\rm NLS}_{L_{2(1)}}\rangle$ 
due to the switch in indices, $\eta=1$ and $2$. 
To explain this further, when approaching along $\vb*{b}_{1}$, 
we find the NLS Bloch state $|\psi^{\rm NLS}_{\vb*{k}^, 1}\rangle$ 
consisting of loops, $|u^{\rm NLS}_{L_{2}}\rangle$, running in direction 2, 
orthogonal to $\vb*{b}_{1}$ ($\vb*{b}_{1}\cdot\vb*{a}_{2} = 0$). 
In other words, the loop $|u^{\rm NLS}_{L_{2}}\rangle$ regarded as a wavefront 
``propagates" in the direction 1 to form $|\psi^{\rm NLS}_{\vb*{k}^*, 1}\rangle$.
Therefore, $\eta$ indexes the direction of the wavefront and not the winding direction of the loop state. 
\par
{\it Degenerate flat bands at $d=3$.}\; 
We can generalize the above arguments to three and higher dimensions.
One possible complication is the degeneracy of flat bands.
Suppose the system has $n_f$ flat bands, then we can choose a set of $n_f$ linearly independent 
normalized Bloch CLS $\{ |\psi^{{\rm CLS}(j)}_{\vb*{k}} \rangle\}$, with $j=1, 2, \cdots n_f$. 
If all these states simultaneously vanish at $\vb*{k}^*$, this momentum is singular. 
In approaching $\vb*{k}^*$ along $\vb*{b}_{\eta}$, each normalized Bloch CLS converges to a corresponding Bloch NLS, 
whose form $|\psi^{{\rm NLS}(j)}_{\vb*{k}^*, \eta}\rangle$, 
depends on the choice of $\{ |\psi^{{\rm CLS}(j)}_{\vb*{k}} \rangle\}$.
However, at least $|\psi^{{\rm NLS}(j)}_{\vb*{k}^*, \eta}\rangle$ must live in the space orthogonal to $\vb*{b}_{\eta}$,  
so that $\eta$ indexes the direction of the wavefront again, i.e. $|\psi^{{\rm NLS}(j)}_{\vb*{k}^*, \eta}\rangle$ is the linear combination of one-loop NLS's in the subspace spanned by the lattice vectors $\vb*{a}_{\mu}$'s, 
which satisfy $\vb*{b}_{\eta}\cdot\vb*{a}_{\mu} = 0$.
\par
As an example, we consider the pyrochlore lattice at $d=3$, which hosts two-fold degenerate flat bands. 
As a counterpart of Eq.~(\ref{eq:bloch-v0}), we can choose two CLSs $|u^{\rm CLS(1)}_{\vb*{k}}\rangle$ and $|u^{\rm CLS(2)}_{\vb*{k}}\rangle$, 
as their basis set, given as 
\begin{eqnarray}
\left\{
\begin{array}{l}
v^{(1)}_{0,\vb*{k}} = e^{-ik_3} - e^{-ik_1}, \\
v^{(1)}_{1,\vb*{k}} = 1 - e^{-ik_3},\\
v^{(1)}_{2,\vb*{k}} = 0,\\
v^{(1)}_{3,\vb*{k}} = e^{-ik_1} - 1,
\end{array}
\right.
\left\{
\begin{array}{l}
v^{(2)}_{0,\vb*{k}} = e^{-ik_3} - e^{-ik_2}, \\
v^{(2)}_{1,\vb*{k}} = 0,\\
v^{(2)}_{2,\vb*{k}} = 1 - e^{-ik_3},\\
v^{(2)}_{3,\vb*{k}} = e^{-ik_2} - 1,
\end{array}
\right.
\label{eq:bloch-pyrochlore}
\end{eqnarray}
each constructed from the primitive CLSs localized around hexagons on the sliced kagome plane, 
$\vb*{a}_1$-$\vb*{a}_3$ and $\vb*{a}_2$-$\vb*{a}_3$, respectively (Fig~\ref{f1} (a)). 
Both vanish at $\vb*{k}=\vb*{k}^*=0$, that marks the singularity. 
For instance, in approaching $\vb*{k}^*$ along $\vb*{b}_3$, 
the linear combination of normalized Bloch CLSs, 
$\alpha|\psi^{\rm CLS (1)}_{\vb*{k}} \rangle + \beta|\psi^{\rm CLS (2)}_{\vb*{k}} \rangle$, 
converges to the superposition of $|u^{\rm NLS}_{L_1}\rangle$ and $|u^{\rm NLS}_{L_2}\rangle$ 
winding along directions 1 and 2, and hence living on the [111]
kagome plane orthogonal to $\vb*{b}_3$~\footnote{See Sec. A in Supplemental Material}.
To see this, it is enough to notice 
$\frac{\partial}{\partial k_3}(\alpha\vb*{v}^{(1)}_{\vb*{k}^*} + \beta\vb*{v}^{(2)}_{\vb*{k}^*})\propto(\alpha + \beta, -\alpha, -\beta, 0)$, from Eq.~(\ref{eq:bloch-pyrochlore}), 
namely the two NLSs are placed on sublattices, $\mu=0,1,2$. 

\begin{figure}[t]
	\begin{center}
		\includegraphics[width=8.5cm]{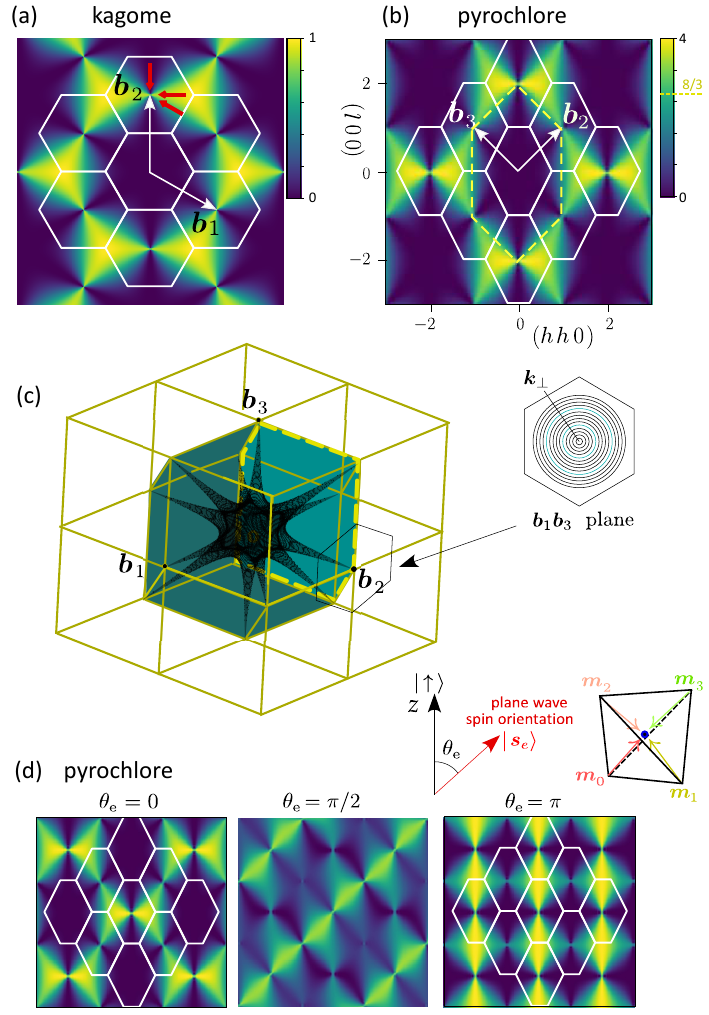}
		\caption{Structure factors of flat bands: 
(a) kagome and (b,c,d) pyrochlore lattice.  
A dashed hexagon in (b) and yellow solid lines in (c) denote the line ${\cal S}(\bm k)=8/3$. 
In panel (c) the dark region ${\cal S}(\bm k)\le 0.01$ is shown near the $\Gamma$-point 
which extends toward 12 different singular points as dark cones, whose slices are shown in the inset. 
Panel (d) shows the case where the removed electron spin varies its angle $\theta_e$ 
about the $z$-axis for the spin-orbit coupled flat bands. Inset shows spin-rotation axes $\vb*{m}_\mu$ 
pointing toward the center of the tetrahedron. 
}
	\label{f2}
	\end{center}
\end{figure}
\par
{\it From NLS to Pinch point.} 
In diffraction measurements, we are concerned with the structure factor, which involves assessing the matrix element containing spatial details regarding sublattice coordinates. 
For example, in photoemission spectroscopy~\cite{RMP_2003,Moser_2017,Day2019,RMP_2021}, we are interested in the operator, 
\begin{equation}   
a^{\dag}_{\vb*{k}} = \frac{1}{\sqrt{N_c}}\sum_{l=0}^{N_c-1}
\sum_{\mu=0}^{n_s-1} e^{i\vb*{k}\cdot(\vb*{R}_l + \vb*{r}_{\mu})}c^{\dag}_{l,\mu},
\label{eq:planewave}
\end{equation} 
where $c^{\dag}_{l,\mu}$ creates an electron in the state $|\vb*{R}_l, \mu\rangle$. 
We now start from the $N_e$ electron ground state, $|\Psi^{N_e} \rangle$, and examine the probability that 
one electron is added to the empty flat band state $|\psi^{\rm CLS}_{\vb*{k}}\rangle$. 
The charge structure factor measured through the photoemission spectroscopy experiment is evaluated by the matrix element, 
$\mathcal{S}(\vb*{k})=|\langle \Psi^{N_e}|c^{\rm CLS}_{\vb*{k}}a^\dag_{\vb*{k}} |\Psi^{N_e} \rangle|^2$, 
which can be rewritten using the normalized Bloch function, Eq.~(\ref{eq:normalizedCLS_detail}) as
\begin{align}
\mathcal{S}(\vb*{k}) &= \bigg|\sum_{\mu} e^{-i{\vb*{k}\cdot{\vb*{r}}_{\mu}} } \: 
\frac{v_{\mu,\bm k}}{|\vb*{v}_{\vb*{k}}|} \bigg|^2 
\xrightarrow[\substack{\vb*{k}\to \vb*{k}^*+\vb*{G}\\ \hspace{-0.6cm}\eta}]{} 
\bigg|\,{\vb*{P}}\cdot \frac{\partial \vb*{v}_{\vb*{k}^*}}{\partial k_{\eta}}\,\bigg|^2, 
\label{eq:sk_general} \\
\vb*{P}&=(e^{-i(\vb*{k}^*+\vb*{G})\cdot {\vb*{r}}_{\mu}}). 
\label{eq:p}
\end{align}
The vanishing of the norm, $|\vb*{v}_{\vb*{k}^*}|=0$, gives the discontinuity of 
$\mathcal{S}(\vb*{k})$ at $\vb*{k}=\vb*{k}^*$, 
implying nothing but the pinch point singularity. 
This singularity potentially appears at a set of momenta, $\vb*{k}^* + \vb*{G}$, 
shifted by arbitrary reciprocal vector $\vb*{G}=\sum_{\mu}g_{\mu}\vb*{b}_{\mu}$, with integers $g_{\mu}$. 
Whereas, the profile of $\mathcal{S}(\vb*{k}\sim \vb*{k}^* + \vb*{G})$ may differ depending on $\vb*{G}$, 
because the phase factor $e^{-i{\vb*{k}}\cdot{\vb*{r}}_{\mu}}$ has different periodicities about $\vb*{G}$
\footnote{The dependence of spectrum on phase factor is systematically studied by Conlon and Chalker
\protect~\cite{Conlon2010} in different context.}. 
\par
Indeed, the profile is determined by the two factors that appear on the r.h.s. of Eq.~(\ref{eq:sk_general}). 
One factor is the vector $\vb*{P}\equiv(e^{-i({\vb*{k}^*}+\vb*{G})\cdot{\vb*{r}}_{\mu}})$ and the other is 
the weight of the NLSs, $(\partial \vb*{v}_{\vb*{k}^*}\big/\partial k_{\eta})$. 
The former depends on the choice of $\vb*{k}^*+\vb*{G}$ 
but not on the approaching direction $\eta$ to $\vb*{k}^*$, 
whereas for the latter the dependence is vice versa. 
The inner product of these two factors governs the profile of $\mathcal{S}(\vb*{k})$, 
indicating that the interference of $d$ independent NLSs 
occurs according to the selection rule encoded in $\vb*{P}$. 
We call the vector $\vb*{P}$ {\it ``a polarizer"} 
as it produces different images by varying the choice of NLSs when $\vb*{k}^*$ is translated by $\vb*{G}$. 
\par
{\it Examples.} 
Revisiting the kagome lattice, we focus on the singular point $\vb*{k} = \vb*{b}_2$, 
where the polarizer takes the form: 
${\vb*{P}} = (e^{-i{\vb*{b}_2}\cdot{\vb*{r}}_{0}}, e^{-i{\vb*{b}_2}\cdot{\vb*{r}}_{1}}, e^{-i{\vb*{b}_2}\cdot{\vb*{r}}_{2}})
=(1, 1, -1)$. 
If we approach $\vb*{k}^*$ along $\vb*{b}_2$ (vertical line in Fig.~\ref{f2}(a)), 
the NLSs are chosen as $\frac{\partial \vb*{v}_{\vb*{k}^*}}{\partial k_{2}}\propto(1, -1, 0)$, 
and $\mathcal{S}(\vb*{k})$ vanishes as $\propto |1\cdot1 + 1\cdot(-1) + (-1)\cdot0|^2=0$. 
Along $\vb*{b}_1$, it yields $\mathcal{S}(\vb*{k}) \sim  2^2=4$. 
When approaching along a general direction, $\vb*{k} \parallel (\alpha\vb*{b}_1+\beta\vb*{b}_2)$, 
only the $\vb*{b}_2$ component contributes according to Eq.~(\ref{eq:NLS_convergence}), 
and $\mathcal{S}(\vb*{k})$ takes a finite value. 
With these simple considerations we can reproduce a bowtie structure. 
\par
Even in the pyrochlore model with degenerate flat bands, 
we can identify the dark line where ${\cal S}(\vb*{k})$ vanishes; 
we may target $\vb*{k}=\vb*{b}_3$ whose polarizer is 
$\vb*{P}=(e^{-i{\vb*{b}_3}\cdot{\vb*{r}}_{0}}, e^{-i{\vb*{b}_3}\cdot{\vb*{r}}_{1}}, e^{-i{\vb*{b}_3}\cdot{\vb*{r}}_{2}}, e^{-i{\vb*{b}_3}\cdot{\vb*{r}}_{3}}) = (1, 1, 1, -1)$. 
Approaching this point parallel to $\vb*{b}_3$, NLSs have weights, 
$\frac{\partial}{\partial k_3}(\alpha\vb*{v}^{(1)}_{\vb*{k}^*} + \beta\vb*{v}^{(2)}_{\vb*{k}^*})\propto(\alpha + \beta, -\alpha, -\beta, 0)$, which is orthogonal to the polarizer and 
${\cal S}(\vb*{k})={\vb*{P}}\cdot \frac{\partial \vb*{v}_{\vb*{k}^*}}{\partial k_{3}}=0$. 
This result is confirmed by the exact analytical form of ${\cal S}(\vb*{k})$ given in Supplemental Material~\footnote{See Sec. B and Eq.~(S16) in Supplemental Material}. 
Figure~\ref{f2}(c) visualizes the low-intensity region, ${\cal S}(\vb*{k})\le 0.01$. 
This region develops around dark lines branching from the $\Gamma$-point toward 12 different singular points, 
where the two dark cones touch. 
\par
{\it Spin-dependent polarizer.}
One caveat is that the function of the polarizer depends on models and experimental probes. 
As an interesting case study, we consider the spin-polarized photoemission spectroscopy applied to the spin-orbit coupled flat band state on the pyrochlore lattice~\cite{nakai2022,nakai2023}.
Suppose we have a half-filled flat band in the ground state, 
with sublattice spins polarized as $|\vb*{s}_{\mu}\rangle\equiv(\vb*{m}_{\mu}\cdot\bm{\sigma})|\uparrow\rangle$, 
where $\vb*{m}_{\mu}$ are the axes shown in Fig.~\ref{f2}(d).
In the inverse photoemission spectroscopy to remove an electron from this ground state, 
the polarizer takes the form, $P = e^{-i({\vb*{k}^*}+\vb*{G})\cdot{\vb*{r}}_{\mu}}\langle\vb*{s}_e|\vb*{s}_{\mu}\rangle$, 
where $|\vb*{s}_{e}\rangle$ stands for the spin state of the removed electron. 
Due to the additional spin-dependent matrix element $\langle\vb*{s}_e|\vb*{s}_{\mu}\rangle$,  
the dark line rotates with the spin orientation of the removed electron, $\theta_{\rm e}$, 
as shown in Fig.~\ref{f2}(d)~\footnote{See Sec. C in Supplemental Material}. 
By carefully preparing a polarizer with dressed matrix elements, the information drawn from the 
spectrum can be used to clarify the microscopic spin structure of the state at focus. 
\par
\begin{figure}[t]
	\begin{center}
		\includegraphics[width=9cm]{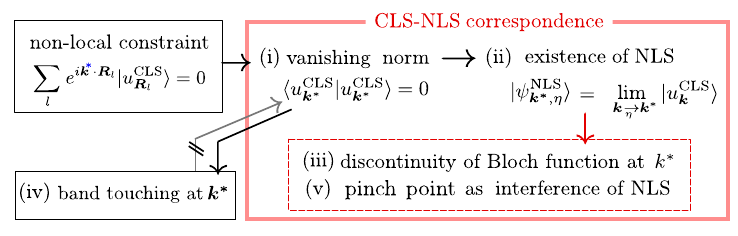}
		\caption{CLS-NLS correspondence and the relationships with the nature of flat bands and pinch points, (i)-(v). 
}
	\label{f3}
	\end{center}
\end{figure}
{\it Summary and Discussion.} 
We have established the CLS-NLS correspondence, identifying the Bloch NLS at the singular $\vb*{k}^*$ point as the momentum derivative of the Bloch CLS. 
The CLS-NLS correspondence makes a transparent connection among the flat band phenomena (i)-(v) as shown in Fig.~\ref{f3}.
Here, (i) the vanishing of the norm of Bloch CLS  enables the operation of momentum derivative in approaching the singular momentum,  from which we reach the form of NLS in (ii)~\footnote{The combination of (i) and (ii) account for (iv) the presence of band touching~\cite{Bergman2008,Rhim2019}. }.
NLS thus obtained is selected from among $d$ dependent choices, depending on the direction in approaching $\vb*{k}^*$ ,  which naturally explains (iii) the discontinuity of the flat band wavefunction.
Finally, the NLS tomography around the singular momentum provides a simple way to describe (v) the structure of pinch point, as an interference pattern of the NLS.

To note, any flat bands including non-singular ones host CLSs, given a general construction~\cite{Rhim2019}, and supported by the arbitrarity of the choice of CLSs, ``NLS" is derived from our formula. However, when CLSs span a complete flat band basis, such ``NLS" is contractible and (ii) and (iii) break down. 
Band touching may occur in that case as well. 
\par
The CLS-NLS correspondence guarantees that we can always have NLS for singular flat bands, 
and can use it for identifying pinch points, or to extract from pinch points the 
electronic or magnetic structures of more complex systems with multi-degrees of freedom. 
\par
Although the system with perfect flat bands and ideal pinch point may seem to be a specific example, 
in reality there are a wide variety of materials with nearly flat bands in its vicinity. 
These classes of bands are allowed to host finite Chern numbers or may give another spectral feature 
known as half moons or ``shadow pinch point"~\cite{Conlon2010,PhysRevB.94.104416,Mizoguchi2017,Mizoguchi2018,Yan2018}. 
The variety of spectral patterns found there can be diagnosed using in a language of pinch points,  
and thus our CLS-NLS correspondence will provide a guiding principle to understand a broad range of systems close
to flat band.
\begin{acknowledgments}
H.N. was supported by a Grant-in-Aid for JSPS Research Fellow (Grant No. JP23KJ0783). 
This work is supported by KAKENHI Grant No. JP20H05655, JP21H05191, JP21K03440 and JP22H01147
from JSPS of Japan.
\end{acknowledgments}

%


\widetext
\pagebreak

\renewcommand{\theequation}{S\arabic{equation}}
\renewcommand{\thefigure}{S\arabic{figure}}
\renewcommand{\thetable}{S\arabic{table}}
\setcounter{equation}{0}
\setcounter{figure}{0}
\setcounter{table}{0}

\begin{center}
\Large 
{Supplemental Material}
\end{center}

%
\subsection{Convention of pyrochlore lattice and the choice of reciprocal lattice vectors}
\label{sec:pyrochlore_convention}
\begin{figure}[h]
   \centering
   \includegraphics[width=8.5cm]{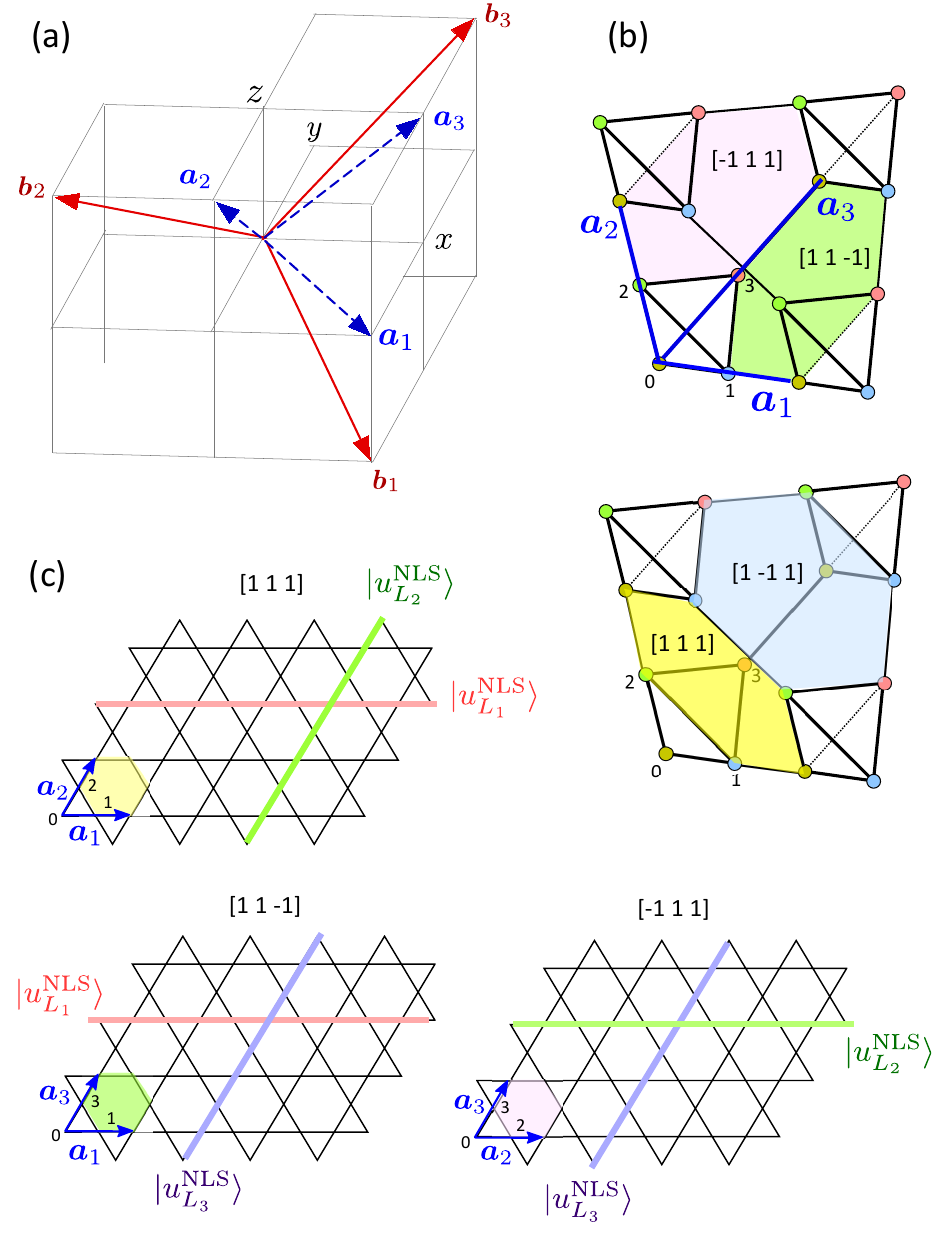}
   \caption{(a) Lattice conventions: the directions of lattice vectors and reciprocal lattice vectors. 
(b) Four hexagons forming a polygon. The hexagons share their vertices and by summing up the four CLS states 
on these hexagons, they cancel out. 
According to the vectors normal to the hexagons we label the surfaces by $[-11\,1], [1\,1-1], [1\,1\,1]$, and $[1-1\,1]$. 
(c) The [1\,1\,1] plane spanned by $\vb*{a}_1$ and $\vb*{a}_2$ supports the NLSs living on the sublattices $\mu=0,1,2$.
}
\label{fS1}
\end{figure}

We take the lattice vectors of the pyrochlore lattice as 
$\vb*{a}_1=(1,-1,0)/2, \vb*{a}_2=(0,-1,1)/2, \vb*{a}_3=(1,0,1)/2$, 
which gives $\vb*{b}_1=(1,-1,-1)2\pi, \vb*{b}_2=(-1,-1,1)2\pi, \vb*{b}_3=(1,1,1)2\pi$ 
shown in Fig.~\ref{fS1}(a). 
In the pyrochlore lattice, we have four independent kagome planes which are denoted as 
$[-1\ 1\,1], [1\,1-\!1], [1\,1\,1]$, and $[1-\!1\,1]$ by the vectors normal to these planes. 
If we construct a hexagonal CLS on the four species of hexagons on these planes, 
the four that share half of the edges and form a polygon are not independent of each other. 
Namely, the summation over them cancels out (see Fig.~\ref{fS1}(b)). 
In this sense, only three of the four CLSs are linearly independent.
\par
By using them we can construct a Bloch CLS confined within one sheet of kagome plane in each direction, as we 
do in the main text Eq.~(4), which are shown schematically in Fig.~\ref{fS1}(c) as colored hexagons.
We can also visualize one-loop NLS on each plane.
When we choose the $[111]$ kagome plane, the choices of three sublattices are $\mu=0,1,2$, and we find two NLSs,
\begin{align}
&|u_1^{\rm NLS}\rangle=\sum_n (|n\vb*{a}_1,0\rangle -|n\vb*{a}_1,1\rangle),\nonumber\\
&|u_2^{\rm NLS}\rangle=\sum_n (-|n\vb*{a}_2,0\rangle +|n\vb*{a}_2,2\rangle),
\end{align}
which constitute the Bloch NLS, when approaching the singular $\vb*{k}^*$ point along $\vb*{b}_3$.
\par
Similarly, for the $[11-1]$ plane, we find one-loop NLSs,
\begin{align}
&|u_1^{\rm NLS}\rangle=\sum_n (|n\vb*{a}_1,0\rangle -|n\vb*{a}_1,1\rangle),\nonumber\\
&|u_3^{\rm NLS}\rangle=\sum_n (-|n\vb*{a}_3,0\rangle +|n\vb*{a}_3,3\rangle),
\end{align}
and for the $[-111]$ plane,
\begin{align}
&|u_2^{\rm NLS}\rangle=\sum_n (|n\vb*{a}_2,0\rangle -|n\vb*{a}_2,2\rangle),\nonumber\\
&|u_3^{\rm NLS}\rangle=\sum_n (-|n\vb*{a}_3,0\rangle +|n\vb*{a}_3,3\rangle). 
\end{align}

%
\subsection{Exact form of the spectral function}
\subsubsection{general}
If two or more flat bands exist, the formula of the structure factor must be generalized.
The expression for the $n_f$-fold degenerate flat bands takes the following form,
 \begin{align}
&S(\vb*{k})=\sum_{\mu, \mu'=0,\cdots n_s-1}S_{\mu,\mu'}(\vb*{k}), \nonumber\\
&S_{\mu,\mu'}(\vb*{k})=P_{\mu}P^*_{\mu'}
   \sum_{m,m'=1,\cdots n_f}v^{(m)}_{\mu,\vb*{k}} (U^{-1}_{\vb*{k}})_{mm'}v^{(m')*}_{\mu',\vb*{k}}. 
    \label{eq:degenerate_Sk}
 \end{align}
 Here, $U(\vb*{k})$ is the Gram matrix composed of the degenerate Bloch CLS bases, 
 which are not necessarily orthogonal or normalized; 
  \begin{equation}
 U_{\vb*{k}}\!=\!\begin{bmatrix}
({\vb*{v}}^{(1)*}_{\vb*{k}}\cdot  {\vb*{v}}^{(1)}_{\vb*{k}}) & ({\vb*{v}}^{(1)*}_{\vb*{k}}\cdot  {\vb*{v}}^{(2)}_{\vb*{k}}) & \cdots &  ({\vb*{v}}^{(1)*}_{\vb*{k}}\cdot  {\vb*{v}}^{(n_f)}_{\vb*{k}})\\
({\vb*{v}}^{(2)*}_{\vb*{k}}\cdot  {\vb*{v}}^{(1)}_{\vb*{k}}) & ({\vb*{v}}^{(2)*}_{\vb*{k}}\cdot  {\vb*{v}}^{(2)}_{\vb*{k}}) & \cdots &  ({\vb*{v}}^{(2)*}_{\vb*{k}}\cdot  {\vb*{v}}^{(n_f)}_{\vb*{k}})\\
 \vdots & \vdots  & \ddots & \vdots\\
({\vb*{v}}^{(n_f)*}_{\vb*{k}}\cdot  {\vb*{v}}^{(1)}_{\vb*{k}}) & ({\vb*{v}}^{(n_f)*}_{\vb*{k}}\cdot  {\vb*{v}}^{(2)}_{\vb*{k}}) & \cdots &  ({\vb*{v}}^{(n_f)*}_{\vb*{k}}\cdot  {\vb*{v}}^{(n_f)}_{\vb*{k}})
 \end{bmatrix}. 
 \label{eq:Gram}
 \end{equation}
$P_{\mu}$ is the sublattice $\mu$ component of the polarizer, and takes the form $P_{\mu}=e^{-i\vb*{k}\cdot\vb*{r}_{\mu}}$ in the simplest case of photoemission spectroscopy, considered in the main text.
If the flat band is not degenerate, the expression Eq.~(\ref{eq:degenerate_Sk}), combined with Eq.~(\ref{eq:Gram}), is reduced to Eq.~(12) in the main text.

\subsubsection{kagome lattice}
Only a single flat band exists for the kagome lattice model. As written as Eq.~(5) in the main text, the Bloch CLS takes the form,
\begin{eqnarray}
\vb*{v}_{\vb*{k}} = 
\begin{bmatrix}
v_{0,\vb*{k}}\\
v_{1,\vb*{k}}\\
v_{2,\vb*{k}}
\end{bmatrix}
=
\begin{bmatrix}
e^{-ik_2} - e^{-ik_1}\\
1 - e^{-ik_2}\\
e^{-ik_1} - 1
\end{bmatrix}
.
\end{eqnarray}
Putting this expression into Eq.~(\ref{eq:degenerate_Sk}), we obtain
\begin{equation}
\mathcal{S}(\vb*{k})=\frac{\Big|\sin\frac{k_1-k_2}{2}\!+\!\sin\frac{k_2}{2}\!-\!\sin\frac{k_1}{2}\Big|^2}
{\sin^2\frac{k_1-k_2}{2}\!+\!\sin^2\frac{k_2}{2}\!+\!\sin^2\frac{k_1}{2}}.
\end{equation}
By choosing the lattice coordinates, $\vb*{a}_1=(1,0)$ and $\vb*{a}_2=(\frac{1}{2},\frac{\sqrt{3}}{2})$, which gives $k_1 = k_x$ and $k_2 = \frac{k_x+\sqrt{3}k_y}{2}$,
we obtain $S(\vb*{k})$ as depicted in Fig.~2(a) in the main text.

\subsubsection{pyrochlore lattice}
A pyrochlore lattice has two-fold degenerate flat bands (except for the spin degeneracy). 
The two basis vectors of Bloch CLS can be chosen as
\begin{eqnarray}
\vb*{v}^{(1)}_{\vb*{k}} = 
\begin{bmatrix}
v^{(1)}_{0,\vb*{k}}\\
v^{(1)}_{1,\vb*{k}}\\
v^{(1)}_{2,\vb*{k}}\\
v^{(1)}_{3,\vb*{k}}
\end{bmatrix}
=
\begin{bmatrix}
e^{-ik_3} - e^{-ik_1}\\
1 - e^{-ik_3}\\
0\\
e^{-ik_1} - 1
\end{bmatrix}
,
\end{eqnarray}
\vspace{0.5cm}
\begin{eqnarray}
\vb*{v}^{(2)}_{\vb*{k}} = 
\begin{bmatrix}
v^{(2)}_{0,\vb*{k}}\\
v^{(2)}_{1,\vb*{k}}\\
v^{(2)}_{2,\vb*{k}}\\
v^{(2)}_{3,\vb*{k}}
\end{bmatrix}
=
\begin{bmatrix}
e^{-ik_3} - e^{-ik_2}\\
0\\
1 - e^{-ik_3}\\
e^{-ik_2} - 1
\end{bmatrix}
.
\end{eqnarray}
$\vb*{v}^{(1)}_{\vb*{k}}$ and $\vb*{v}^{(2)}_{\vb*{k}}$ are respectively constructed from the primitive CLSs localized on kagome planes, $\vb*{a}_1-\vb*{a}_3$ and $\vb*{a}_2-\vb*{a}_3$, as defined in Sec.~\ref{sec:pyrochlore_convention}.

By inputting them into Eqs.~(\ref{eq:degenerate_Sk}) and (\ref{eq:Gram}), we find the diagonal components of $ S_{\mu,\mu'}(\vb*{k})$ satisfy
\begin{eqnarray}
 \sum_{\mu=0,1,2,3}S_{\mu,\mu}(\vb*{k}) = 2.
 \end{eqnarray}
For the off-diagonal components, $\mu\not=\mu'$, we obtain
 \begin{eqnarray}
 S_{\mu,\mu'}(\vb*{k})
 =\Omega_{\vb*{k}}^{-1}\bigg[ 
 \sum_{\mu''=0,1,2,3}\sin\frac{k_{\mu} - k_{\mu''}}{2}\sin\frac{k_{\mu''} - k_{\mu'}}{2}
 \bigg]^{-1},\nonumber\\
 \end{eqnarray}
 where $k_0=0$, and
 \begin{eqnarray}
 \Omega_{\vb*{k}} 
&=& \sin^2\frac{k_1}{2} + \sin^2\frac{k_2}{2} + \sin^2\frac{k_3}{2} + \sin^2\frac{k_1-k_2}{2} + \sin^2\frac{k_1-k_3}{2} + \sin^2\frac{k_2-k_3}{2}, 
 \label{eq:Sq_denominator}
 \end{eqnarray}
 and each component is given as 
 \begin{eqnarray}
 &&S_{01}(\vb*{k})=\Omega_{\vb*{k}}^{-1} \bigl[\sin\frac{k_2}{2}\sin\frac{k_1-k_2}{2} + \sin\frac{k_3}{2}\sin\frac{k_1-k_3}{2}\bigr],  
 \nonumber\\ 
 &&S_{02}(\vb*{k})=\Omega_{\vb*{k}}^{-1} \bigl[-\sin\frac{k_1}{2}\sin\frac{k_1-k_2}{2}+ \sin\frac{k_3}{2}\sin\frac{k_2-k_3}{2}\bigr], 
 \nonumber\\ 
 &&S_{03}(\vb*{k})=\Omega_{\vb*{k}}^{-1} \bigl[-\sin\frac{k_1}{2}\sin\frac{k_1-k_3}{2} - \sin\frac{k_2}{2}\sin\frac{k_2-k_3}{2}\bigr], 
 \nonumber\\
 &&S_{12}(\vb*{k})=\Omega_{\vb*{k}}^{-1} \bigl[-\sin\frac{k_1}{2}\sin\frac{k_2}{2} - \sin\frac{k_1-k_3}{2}\sin\frac{k_2-k_3}{2}\bigr], 
 \nonumber \\
 &&S_{13}(\vb*{k})=\Omega_{\vb*{k}}^{-1} \bigl[-\sin\frac{k_1}{2}\sin\frac{k_3}{2} + \sin\frac{k_1-k_2}{2}\sin\frac{k_2-k_3}{2}\bigr], 
 \nonumber\\ 
 &&S_{23}(\vb*{k})=\Omega_{\vb*{k}}^{-1} \bigl[-\sin\frac{k_2}{2}\sin\frac{k_3}{2} - \sin\frac{k_1-k_2}{2}\sin\frac{k_1-k_3}{2}\bigr]. 
 \nonumber\\
 \label{eq:Sqs}
 \end{eqnarray}
 The other components are given as $ S_{\mu,\mu'}(\vb*{k})= S_{\mu',\mu}(\vb*{k})$. 
 We finally obtain the exact form of the structure factor as 
  \begin{eqnarray}
  S(\vb*{k}) &=& 2(1 + S_{01}(\vb*{k}) + S_{02}(\vb*{k})+ S_{03}(\vb*{k}) + S_{12}(\vb*{k}) + S_{13}(\vb*{k}) + S_{23}(\vb*{k})).
   \label{eq:Sqtotal}
  \end{eqnarray}
 
For a reciprocal vector $\vb*{k}=(k_1\vb*{b}_1+k_2\vb*{b}_2+k_3\vb*{b}_3)/2\pi$, 
with $k_1=(k_x-k_y)/2$, $k_2=(-k_y+k_z)/2$, and $k_3=(k_x+k_z)/2$.
Putting them into Eqs.~(\ref{eq:Sq_denominator}), (\ref{eq:Sqs}) and (\ref{eq:Sqtotal}), we obtain the structure factor in a symmetric form,
 \begin{eqnarray}
 \Omega_{\vb*{k}}=\sin^2{\frac{k_x+k_y}{4}} + \sin^2{\frac{k_y+k_z}{4}} + \sin^2{\frac{k_z+k_x}{4}} + \sin^2{\frac{k_x-k_y}{4}} + \sin^2{\frac{k_y-k_z}{4}} + \sin^2{\frac{k_z-k_x}{4}},
 \end{eqnarray}
  \begin{align}
 S(\vb*{k}) = \frac{2}{ \Omega_{\vb*{k}}}\Bigl[1 - &\Bigl(\sin{\frac{k_x-k_y}{4}}\sin{\frac{k_x-k_z}{4}} + \sin{\frac{k_y-k_x}{4}}\sin{\frac{k_y-k_z}{4}} + \sin{\frac{k_z-k_x}{4}}\sin{\frac{k_z-k_y}{4}}\nonumber\\
 &+ \sin{\frac{k_x+k_y}{4}}\sin{\frac{k_x+k_z}{4}} + \sin{\frac{k_y+k_x}{4}}\sin{\frac{k_y+k_z}{4}} + \sin{\frac{k_z+k_x}{4}}\sin{\frac{k_z+k_y}{4}}\Bigr)\nonumber\\
 &+\sin{\frac{k_x+k_y}{4}}\Bigl(\sin{\frac{k_z-k_x}{4}} + \sin{\frac{k_z-k_y}{4}}\Bigr) + \sin{\frac{k_y+k_z}{4}}\Bigl(\sin{\frac{k_x-k_y}{4}} + \sin{\frac{k_x-k_z}{4}}\Bigr)\nonumber\\
 &+ \sin{\frac{k_z+k_x}{4}}\Bigl(\sin{\frac{k_y-k_x}{4}} + \sin{\frac{k_y-k_z}{4}}\Bigr)\Bigr].
 \end{align}
From these expressions, we can obtain the plots in Fig.~2(b) and (c) in the main text.
\par
 
\subsection{Spin-orbit coupled flat band states of the pyrochlore lattice}
We introduce the spin-orbit coupled flat band model on a pyrochlore lattice~\cite{nakai2022,nakai2023}, which was originally introduced to address the exotic low-temperature insulating phase of CsW$_2$O$_6$~\cite{Okamoto:2020aa}.
The tight-binding part of the model contains two parameters, spin-orbit coupling (SOC) $\lambda$ and a standard transfer integral $t$, and the Hamiltonian is written as
\begin{eqnarray}
\label{eq:Ham_kin} 
\mathcal{H}_{\rm kin} = -t_{\rm eff} \sum_{\langle i,j \rangle} \vb*{c}^\dagger_i U_{ij}(\theta) \vb*{c}_j, 
\end{eqnarray}
with $\vb*{c}_i = ( c_{i\uparrow}, c_{i\downarrow} )^{\rm T}$ 
and $t_{\rm eff}=\sqrt{t^2 +\lambda^2}$. 
An SU(2) matrix $U_{ij}(\theta) = e^{-i\theta\hat{\vb*{\nu}}_{ij}\cdot\vb*{\sigma} /2}$ 
rotates the electron spin by angle $\theta=2\arctan(\lambda/t)$ about the $\hat{\vb*{\nu}}_{ij}$-axis 
when it hops from  site $j$ to $i$, and this angle varies with $\lambda/t$. 
Here, we take $t_{\rm eff}=\sqrt{t^2 +\lambda^2}=1$ as an energy unit. 
When  $\theta=-2\arctan (2\sqrt{2})$ we have SOC flat bands. 
\par
In the SOC flat band state, the relative angle of spin orientation of the four sublattices is quenched 
while totally the SU(2) symmetry is kept.  
These spin configurations are given by setting a global spinor state 
$\vb*{\chi}$ and applying a set of operator $\Gamma_\mu$, 
which is the $\pi$ rotation of the spins about the axis $\bm m_\mu$ 
that points from the sublattice to the center of the tetrahedron, 
given as 
\begin{align}
&\bm m_0=(1,-1,1)/2,\quad 
\bm m_1=(-1,1,1)/2,\nonumber \\
&\bm m_2=(1,1,-1)/2,\quad 
\bm m_{3}=(-1,-1,-1)/2. 
\end{align}
When we set $\vb*{\chi}=(1,0)^T$ pointing in the $z$-direction of the pyrochlore $xyz$-axis, 
the orientation of spins on four sublattices yields, 
\begin{align}
 |\vb*{s}_\mu\rangle&=(\cos(\alpha_\mu/2), e^{i\phi_\mu/2}\sin(\alpha_\mu/2))^T, \notag\\ 
 & (\alpha_\mu,\phi_\mu)=\left\{
 \begin{array}{ll}
   (2\pi/3,-\pi/4) & \mu=0 \\
   (2\pi/3,3\pi/4) & \mu=1 \\
   (\pi/3,\pi/4)  & \mu=2 \\
   (\pi/3,5\pi/4) & \mu=3 \\
\end{array}
\right., 
\end{align}
where $\alpha_\mu$ and $\phi_\mu$ are the zenithal and azimuthal angles of spins. 
\par
If finite Hubbard interaction is considered, the $N_e$-electron ground state $|\Psi^{N_e} \rangle$ is still exactly obtained at quarter-filling for $\theta=-2\arctan (2\sqrt{2})$, where the lowest flat band is half occupied. The ground state can be written as
\begin{eqnarray}
|\Psi^{N_e} \rangle = \prod_{m\in{\rm FB}}\prod_{\vb*{k}}c^{\dag}_{\vb*{k},m,\vb*{s}_\mu}|0\rangle,
\end{eqnarray} 
where
\begin{align}
c^{\dag}_{\vb*{k},m,\vb*{s}_\mu} = \frac{1}{\sqrt{N_c}}\sum_{l=0}^{N_c-1}&e^{i\vb*{k}\cdot\vb*{R}_l}
\sum_{\mu=0}^{n_s-1}\frac{\tilde{v}^{(m)}_{\mu,\bm k}}{|\vb*{\tilde{v}}^{(m)}_{\vb*{k}}|}\sum_{\sigma=\uparrow,\downarrow}c^{\dag}_{l,\mu,\sigma}\langle\sigma|\vb*{s}_\mu\rangle.
\label{eq:flatband_creation}
\end{align} 
$c^{\dag}_{\vb*{k},m,\vb*{s}_\mu}$ creates an electron at the $m$-th flat band with the spin state $|\vb*{s}_\mu\rangle$.
Here, the coefficients of flat band Bloch function, $\vb*{\tilde{v}}^{(m)}_{\vb*{k}}$ are chosen to be orthogonal to each other.
\par
In the inverse spin-polarized photoemission spectroscopy, we take out one electron at the spin state $|\vb*{s}_e\rangle$.
The corresponding matrix element is evaluated as
\begin{eqnarray}
\mathcal{S}(\vb*{k})=\sum_{m\in{\rm FB}}|\langle \Psi^{N_e}|a^\dag_{\vb*{k}} c_{\vb*{k},m,\vb*{s}_\mu}|\Psi^{N_e} \rangle|^2, 
\end{eqnarray} 
where the creation operator of plane wave state is written as
\begin{equation}   
a^{\dag}_{\vb*{k}} = \frac{1}{\sqrt{N_c}}\sum_{l=0}^{N_c-1}
\sum_{\mu=0}^{n_s-1} e^{i\vb*{k}\cdot(\vb*{R}_l + \vb*{r}_{\mu})}\sum_{\sigma=\uparrow,\downarrow}c^{\dag}_{l,\mu,\sigma}\langle\sigma|\vb*{s}_e\rangle.
\label{eq:planewave}
\end{equation}   
From Eqs.~(\ref{eq:flatband_creation}) and (\ref{eq:planewave}), we find
\begin{eqnarray}
\langle \Psi^{N_e}|c^{\dag}_{\vb*{k},m,\vb*{s}_\mu} a_{\vb*{k}}|\Psi^{N_e} \rangle=e^{-i\vb*{k}\cdot\vb*{r}_{\mu}}\langle\vb*{s}_e|\vb*{s}_\mu\rangle\sum_{m\in{\rm FB}}\frac{\tilde{v}^{(m)}_{\mu,\bm k}}{|\vb*{\tilde{v}}^{(m)}_{\vb*{k}}|}.
\end{eqnarray} 
The prefactor $P_{\mu}=e^{-i\vb*{k}\cdot\vb*{r}_{\mu}}\langle\vb*{s}_e|\vb*{s}_\mu\rangle$ accounts for the polarizer in this case.
We define the zenithal and azimuthal angles of incident spin $\sigma$ as 
$(\theta_{\rm e},\phi_{\rm e})$ and set $\phi_{\rm e}=0$, 
and obtain the spectrum shown in Fig.~2(d) in the main text for $\theta_{\rm e}=0,\pi/2$, and $\pi$.

\end{document}